\documentclass[aps,amsmath,amsfonts,11pt]{revtex4}
\usepackage{graphicx}
\usepackage{epstopdf}
\usepackage{epsfig,graphicx}
\usepackage[english]{babel}
\usepackage{amsfonts}
\usepackage{amsmath}
\usepackage{latexsym}
\usepackage{graphics,bm}
\usepackage{dcolumn}
\usepackage{bm}
\usepackage{rotating}

\begin{document}

\title{Optomechanical control of mode conversion in a hybrid semiconductor microcavity containing a quantum dot}

\author{ Shahnoor Ali$^{1}$ and Aranya B Bhattacherjee$^{2}$ }

\address{$^{1}$School of Physical Sciences, Jawaharlal Nehru University, New Delhi-110067, India} 
\address{$^{2}$Department of Physics, Birla Institute of Technology and Science, Pilani,
Hyderabad Campus,  Telangana State - 500078, India}

\begin{abstract}
The future of quantum communication systems and quantum networks heavily rely on the ability to fabricate and coherently control the conversion of photons between different modes based on a solid-state plateform. In this work, we propose and theoretically investigate a scheme to optomechanically control coherent mode conversion of optical photons by utilizing two optically coupled hybrid semiconductor microcavities containing a quantum dot (QD). One of the microcavity is pumped by an external laser and the second cavity is driven by light emitted by the QD that is embedded in the interface separating the two microcavities. The semiconductor microcavities can be fabricated using distributed Bragg reflectors (DBR) and can be made movable. We have demonstrated that photon-mode-conversion efficiency can be coherently manipulated by the optomechanical mode under appropriate conditions.
\end{abstract}

\maketitle

\section{Introduction}

Quantum electrodynamics of semiconductor  micro-cavity investigates light-matter interactions inside meso/nanoscale structures. These hybrid systems have emerged as robust and scalable platforms for implementing state of the art of technologies for quantum optics, quantum communications and quantum networks \citep{mabuchi, sri, kim}. In a similar context the ability to coherently manipulate optical mode-conversion in hybrid quantum systems has brought technological implications for both classical communication systems as well as quantum networks \citep{kim, kiel, ritt}. One of the essential requirements of hybrid quantum networks is a low loss interface that is capable of sustaining quantum coherence when two spatially separate systems operating at different frequencies are connected \citep{wall}.

Semiconductor quantum dots (QDs) confined in micro-cavities, due to their high tunability and large density of states are emerging as promising candidates for developing  hybrid quantum devices \citep{vah1, vah2, khit, tang, arka1, arka2, ali, sonam, aranya}. In particular photonic crystal micro cavities have opened avenues due to its ultra high quality factor and highly confined extremely small mode volume. The remarkable feature of solid state based cavity QED is that they can be easily fabricated and integrated  into  large  scale arrays for quantum networks and quantum information  processing applications.

In an optomechanical cavity the optical field can couple to the mechanical oscillator via radiation pressure \citep{kipp, aspel}. Such optomechanical systems has emerged  as a new type of light matter  hybrid quantum interface \citep{stann, tian}. Controlling system with several mechanical modes at the quantum level is a challenge  and efforts are  underway to coherently manipulate them \citep{lin, nog, shk}. These efforts have led to the experimental realization of hybridization and coherent swapping in optomechanical systems \citep{shk, fang, speth, weav}. In the the context of mode conversions, until recently experiments have utilized the intrinsic optical non-linearities of materials \citep{huang, toul, tan, rakher1, rakher2, guinn}. Optomechanical mode conversion  utilising simple hybrid optomechanical systems has been successfully demonstrated \citep{dong, hill}. Optical mode conversion platform based on  a QD embedded in a photonic crystal micro cavity has also been proposed \citep{li}.

In view of the interesting developments that have taken place in semiconductor and optomechanical systems over the past few years in the direction of mode conversion technology, in this aricle, we  investigate a proposal to implement a tunable hybrid optomechanical semiconductor microcavity system for optical mode conversion. The system that we propose consist of  two optically coupled hybrid semiconductor microcavities containing a quantum dot (QD), embedded in the interface separating the two microcavities. The mechanical mode in the system appears in the form of movable distributed Bragg reflectors (DBR) which is utilized to fabricate the semiconductor microcavities. We have demonstrated that optical mode conversion can be tuned and controlled by the optomechanical mode of the DBR under appropriate system parameters.

\section{Proposed model}

The concept of the proposed scheme is illustrated in Fig.1. In the present model, a hybrid double cavity optomechanical system is considered in which a single semiconductor quantum dot (QD) with two energy levels is embedded in the center of the double micro-cavity such that it couples to both the micro-cavity optical modes 1 and 2. Each micro-cavity of the hybrid system supports a single field mode. Here, the micro-cavity is formed by a set of distributed Bragg reflectors (DBR). Along the long axis of the system, light can be confined by the DBR while air guiding dielectric can be used to provide confinement along the transverse direction \citep{gudat}. DBR mirrors are constructed using quarter-wavelength thick high and low refractive index layers with tunable reflectivity \citep{choy}. The coupling of light in/out of the system is also achieved by these DBR. In order to introduce optomechanics into the system, the two set of DBR on the left and the right of the hybrid system can be made movable according to known techniques \citep{choy, yama, bohm}. A CW laser drives the cavity mode 1 as shown in Fig.1. Cavity mode 2 is not directly coupled to the input pump/cavity mode 1. The only way cavity mode 2 can get light is by means of the QD. 
A CW driving laser field $E_{in}(t)=A_{in} e^{-i \omega_{p} t}$ with the driving frequency $\omega_{p}$ and amplitude $A_{in}$ pumps the cavity 1. The two energy levels transition $| 1 \rangle$ $ \leftrightarrow $ $ | 2 \rangle$ of the semiconductor QD is simultaneously coupled to the two optical cavity modes 1 and 2 with resonance frequencies $\omega_{1}$ and $\omega_{2}$ and coupling strengths $g_{1}$, $g_{2}$ respectively. We take $\omega_{qd}$ as the QD optical transition between the ground state $| 1 \rangle$ and the excited state $| 2 \rangle$. The two movable DBR which forms a part of cavity 1 and 2 are assumed to be identical, having similar mechanical frequency $\omega_{m}$. The optical modes of cavity 1,2 are coupled to their respective mechanical DBR with optomechanical coupling strengths $\Omega_{1}$ and $\Omega_{2}$. We will analyze the system by first taking $\Omega_{2}=0$ (i.e DBR of the cavity 2 as stationary) and in second case $\Omega_{1}=0$ (DBR of the cavity 1 as stationary).

\begin{figure}[ht]
\hspace{-0.7cm}
\includegraphics [scale=0.60]{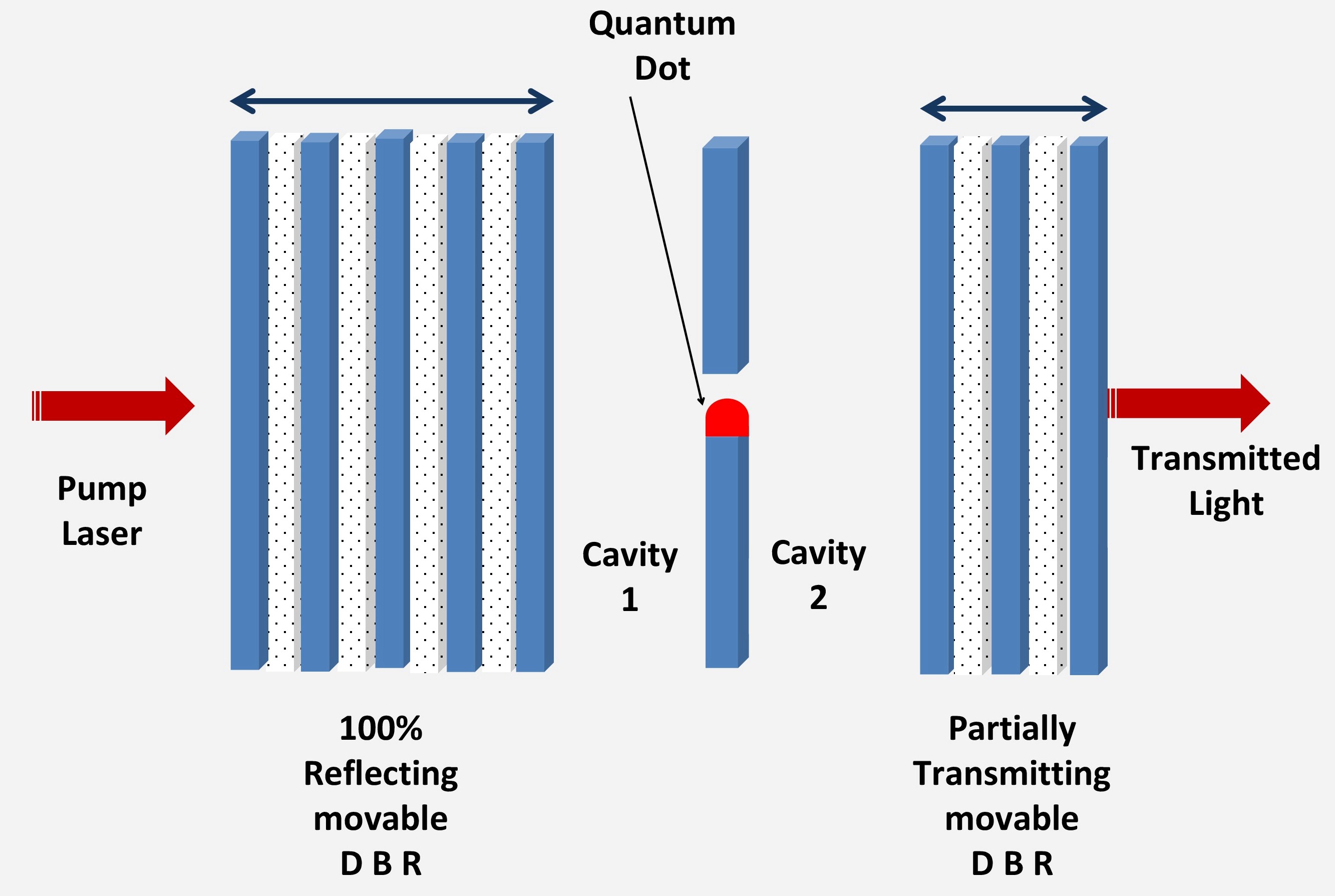} 
\caption{(Color online) Schematic structure of the optical setup studied. It is composed of two semiconductor microcavities fabricated using DBR. Both the DBR's of cavity 1 and 2 can be made movable. The quantum dot is grown on the interface separating the two cavities as shown and hence couples to both the optical modes of cavity 1 and 2. The blue and white strips could correspond to AlGaAs and GaAs layers respectively.  } 
\label{fig:1}
\end{figure}

The linearized optomechanical Hamiltonian describing the hybrid system for the first case ($\Omega_{2}=0$) is given by

\begin{eqnarray}
H_{1} &=& \hbar \Delta_{d} \hat{\sigma}^{\dagger} \hat{\sigma}+\hbar \Delta_{1} \hat{a}_{1}^{\dagger} \hat{a}_{1}+\hbar \Delta_{2} \hat{a}_{2}^{\dagger} \hat{a}_{2}+\hbar \omega_{m} \hat{b}^{\dagger} \hat{b}+\hbar g_{1}(\hat{a}_{1} \hat{\sigma}^{\dagger}+ \hat{a}_{1}^{\dagger} \hat{\sigma}) \nonumber \\
&+& \hbar g_{2}(\hat{a}_{2} \hat{\sigma}^{\dagger}+ \hat{a}_{2}^{\dagger} \hat{\sigma})-\hbar \Omega_{1} (\hat{a}_{1}^{\dagger} \hat{b}+\hat{a}_{1} \hat{b}^{\dagger}) +\hbar \sqrt{\kappa_{1}^{ext}} (A_{in}^{*} \hat{a}_{1}+A_{in} \hat{a}_{1}^{\dagger}) .
\end{eqnarray}

Here the rotating wave approximation (RWA) and the electric-dipole approximation has been used. In deriving the above Hamiltonian 1, we have used the transformation into the rotating frame at frequency $\omega_{p}$ of the driving laser. In the above Hamiltonian 1, the first term is the unperturbed part of the two-level QD with $\Delta_{d}=\omega_{qd}-\omega_{p}$ is the detuning of the QD resonant frequency $\omega_{qd}$ from the driving laser frequency $\omega_{p}$. The second and third terms are the energies of the bare cavity modes 1 and 2 with detuning $\Delta_{i}=\omega_{i}-\omega_{p}$ $(i=1,2)$. The fourth term is the mechanical mode energy with frequency $\omega_{m}$ of the DBR of cavity 1. The fifth and sixth terms are the coherent interaction of the QD with the cavity modes 1 and 2 respectively. The seventh term is the optomechanical interaction of the cavity mode 1 with the movable DBR of cavity 1. Finally the last term describes the driving of the cavity 1. Also $\hat{\sigma}^{\dagger}$ and $\hat{\sigma}$ are the usual Pauli operators describing the electronic transitions of the QD. Further $\hat{a_{1}}^{\dagger}(\hat{a_{1}})$, $\hat{a_{2}}^{\dagger} (\hat{a_{2}})$ and $\hat{b}^{\dagger} (\hat{b})$ describes the creation(destruction) operators for the cavity mode1,2 and mechanical mode respectively. The parameter $\kappa_{1}^{ext}$ is the coupling rate between the optical mode of cavity 1 and the external pumping mode. The corresponding linearized optomechanical Hamiltonian of the system when $\Omega_{2} \neq 0$ and $\Omega_{1}=0$ is

\begin{eqnarray}
H_{2} &=& \hbar \Delta_{d} \hat{\sigma}^{\dagger} \hat{\sigma}+\hbar \Delta_{1} \hat{a}_{1}^{\dagger} \hat{a}_{1}+\hbar \Delta_{2} \hat{a}_{2}^{\dagger} \hat{a}_{2}+\hbar \omega_{m} \hat{b}^{\dagger} \hat{b}+\hbar g_{1}(\hat{a}_{1} \hat{\sigma}^{\dagger}+ \hat{a}_{1}^{\dagger} \hat{\sigma}) \nonumber \\
&+& \hbar g_{2}(\hat{a}_{2} \hat{\sigma}^{\dagger}+ \hat{a}_{2}^{\dagger} \hat{\sigma})-\hbar \Omega_{2} (\hat{a}_{2}^{\dagger} \hat{b}+\hat{a}_{2} \hat{b}^{\dagger}) +\hbar \sqrt{\kappa_{1}^{ext}} (A_{in}^{*} \hat{a}_{1}+A_{in} \hat{a}_{1}^{\dagger}).
\end{eqnarray}

In order to proceed further, we now derive the corresponding quantum-Heisenberg-Langevin equations from the Hamiltonians of Eqns.(1) and (2) using the formalism $\frac{d\hat{O}}{dt}=\frac{1}{i \hbar}[\hat{O},H_{j}] (j=1,2)$. This yields from the Hamiltonian 1,

\begin{equation}
\frac{d \hat{a}_{1}}{dt}=-\left ( i \Delta_{1}+  \frac{\kappa_{1}}{2} \right) \hat{a}_{1}-i g_{1} \hat{\sigma}+i \Omega_{1} \hat{b}-i \sqrt{\kappa_{1}^{ext}}A_{in}+\hat{f}_{a_{1}},
\end{equation}

\begin{equation}
\frac{d \hat{a}_{2}}{dt}=-\left ( i \Delta_{2}+  \frac{\kappa_{2}}{2} \right) \hat{a}_{2}-i g_{2} \hat{\sigma}+\hat{f}_{a_{2}},
\end{equation}

\begin{equation}
\frac{d \hat{b}}{dt}=-\left ( i \omega_{m}+  \frac{\gamma_{m}}{2} \right) \hat{b}+i \Omega_{1} \hat{a}_{1}+\hat{f}_{b},
\end{equation}

\begin{equation}
\frac{d \hat{\sigma}}{dt}=-\left ( i \Delta_{d}+  \frac{\gamma_{qd}}{2} \right) \hat{\sigma}-i g_{1} \hat{a}_{1}-i g_{2} \hat{a}_{2}+\hat{f}_{\sigma}.
\end{equation}

Similar equations can be derived from the Hamiltonian 2. In the above Eqns. (3)-(6), $\kappa_{i} (i=1,2)$ is the total cavity decay rate, $\gamma_{m}$ is the mechanical damping rate while $\gamma_{qd}$ is the QD total decay rate, $\gamma_{qd}=\gamma_{spon}+\gamma_{dep}$, where $\gamma_{spon}$ is the spontaneous emission decay rate and $\gamma_{dep}$ is the dephasing rate. The operators $\hat{f}_{a_{1}}$, $\hat{f}_{a_{2}}$, $\hat{f}_{b}$ and $\hat{f}_{\sigma}$ are the quantum noise operators with  $ \langle \hat{f}_{a_{1}} \rangle = 0$, $ \langle \hat{f}_{a_{2}} \rangle = 0$, $ \langle \hat{f}_{b} \rangle = 0$ and $\langle \hat{f}_{\sigma} \rangle = 0$. We will neglect the expectation values of all the noise operators assuming cold reservoir. The total cavity decay rate $\kappa_{i}=\kappa_{i}^{int}+\kappa_{i}^{ext}$, where $\kappa_{1}^{int}$ is the intrinsic cavity decay rate and $\kappa_{1}^{ext}$ is the effective cavity output coupling rate of mode $i$.

\section{Emission power from cavity modes}

\begin{figure}[ht]
\hspace{-0.7cm}
\begin{tabular}{cc}
\includegraphics [scale=0.60]{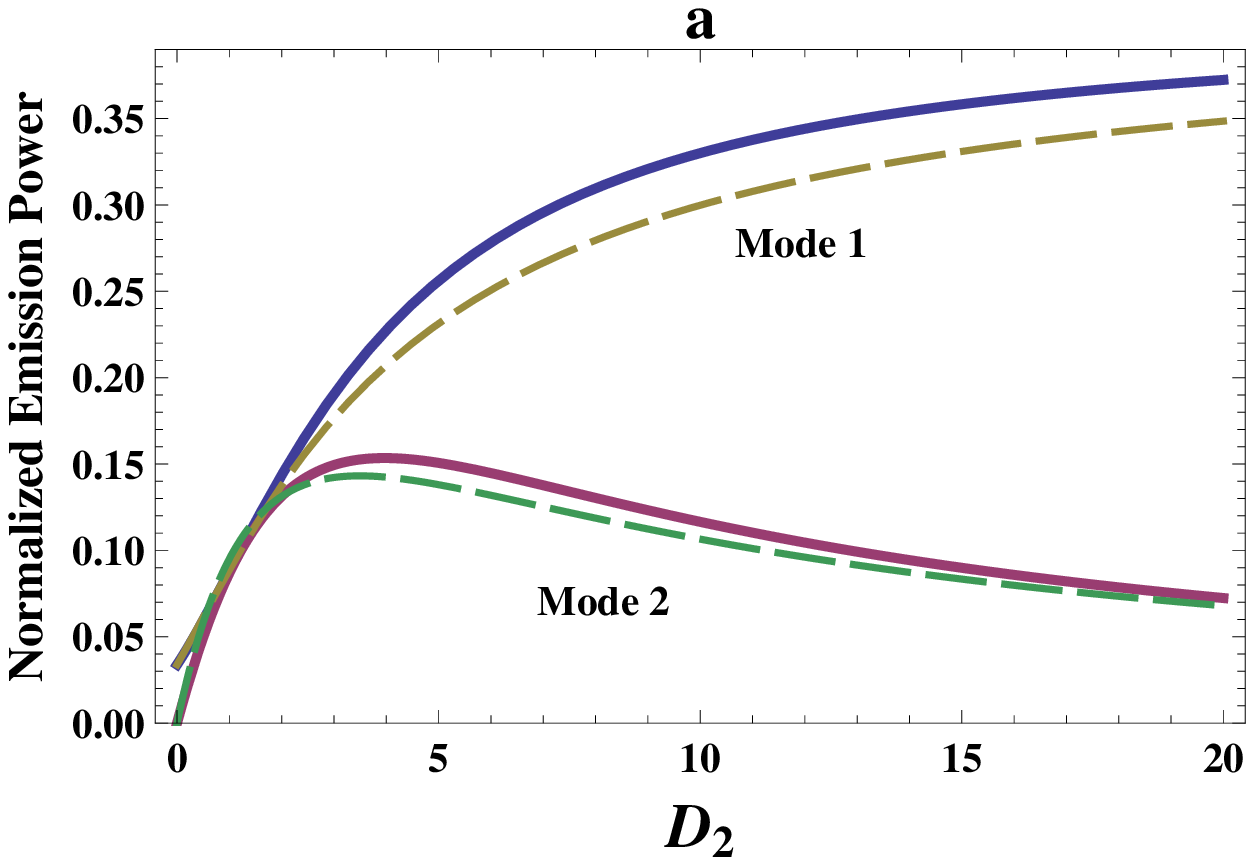} \includegraphics [scale=0.60] {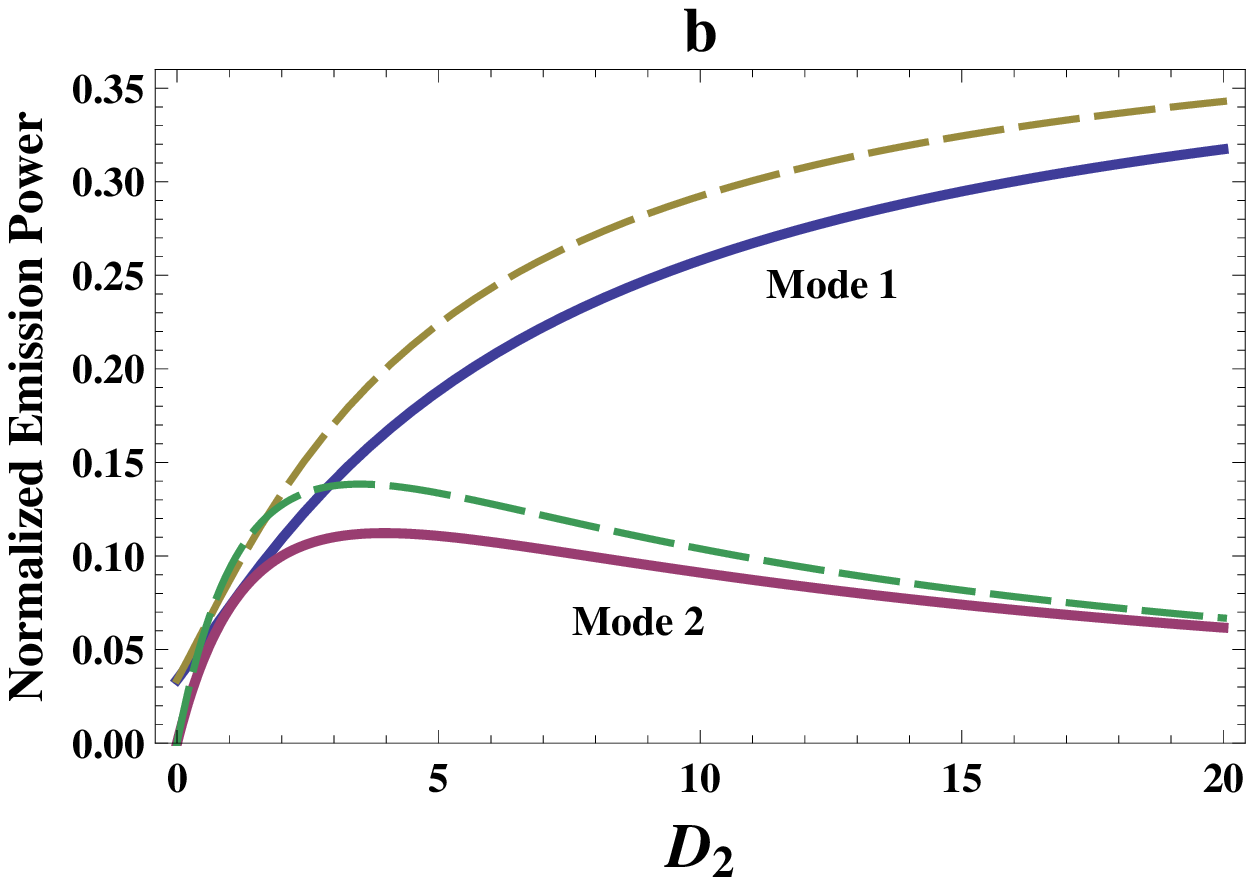}\\
\includegraphics [scale=0.60]{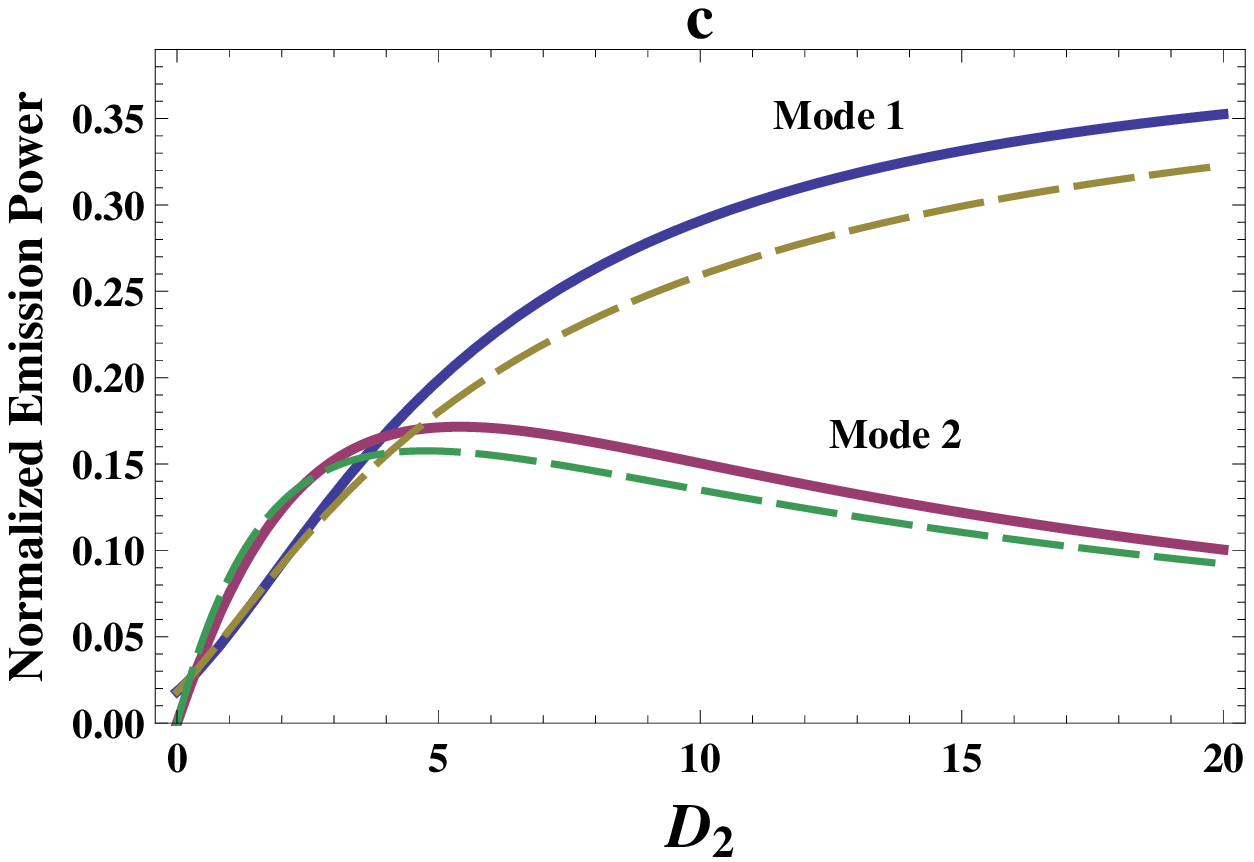} \includegraphics [scale=0.60] {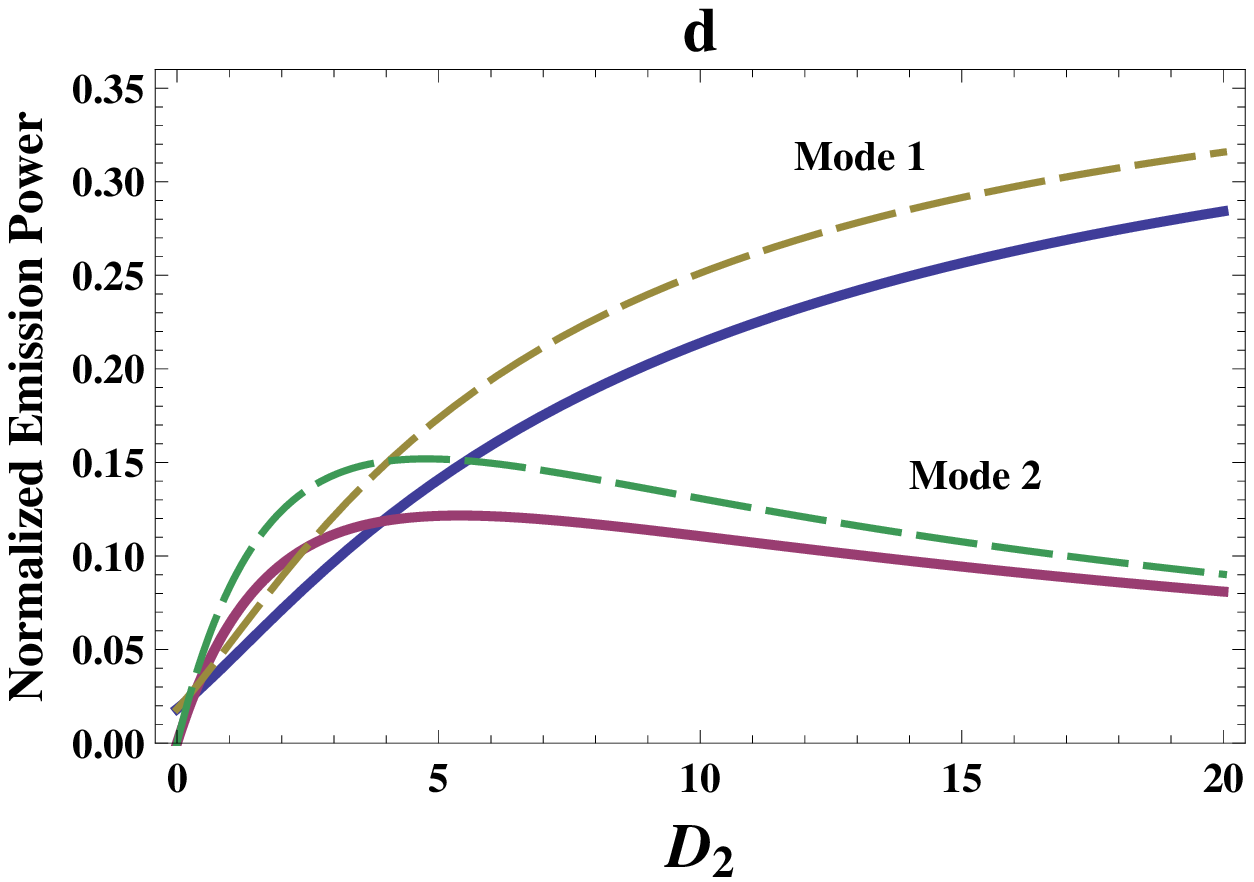}\\
 \end{tabular}
\caption{(Color online)  Normalized optical intracavity emission from mode 1 and mode 2 as a function of cavity cooperativity $D_{2}$ for the case $\Omega_{2} \neq 0$, $\Omega_{1}=0$ and for two different values of optomechanical cooperativity $C_{2}=1$ (dashed line), $C_{2}=10$ (solid line). (a) $D_{1}=4.3$, $\Delta_{1}=0.6 \kappa$, (b) $D_{1}=4.3$, $\Delta_{1}=-0.6 \kappa$, (c) $D_{1}=6.3$, $\Delta_{1}=0.6 \kappa$ and (d) $D_{1}=6.3$, $\Delta_{1}= -0.6 \kappa$. The other system parameters are chosen as $\gamma_{qd}=0.3 \kappa$, $\gamma_{m}=0.001 \kappa$, $\omega_{m}=0.01 \kappa$ and we have taken $\kappa_{1}=\kappa_{2}=\kappa$.   } 
\label{fig:2}
\end{figure}

The operators can be reduced to their expectation values i.e. $\langle \hat{a}_{1}(t)\rangle = a_{1}(t)$, $\langle \hat{a}_{2}(t)\rangle = a_{2}(t)$, $\langle \hat{b}(t)\rangle = b(t)$ and $\langle \hat{\sigma}(t)\rangle = \sigma (t)$ when we are only interested in their mean response \citep{waks, kwon}. In this limit, the expectation values of the noise operators also vanish. The mean-field steady state solutions of Eqns. (3)-(6) for the intracavity field amplitudes of the two optical modes $a_{1s}$ and $a_{2s}$ are respectively, given by

\begin{equation}
|a_{1s}|^{2}=\dfrac{\kappa_{1}^{ext} |A_{in}|^{2} (\omega_{m}^{2}+\frac{\gamma_{m}^{2}}{4}) (\beta_{1R}^{2}+\beta_{1I}^{2}) }{(\alpha_{1R} \beta_{1R}-\alpha_{1I} \beta_{1I}+\gamma_{1R})^{2}+(\alpha_{1R} \beta_{1I}+\beta_{1R} \alpha_{1I}+\gamma_{1I})^{2}},
\end{equation}

\begin{equation}
|a_{2s}|^{2}=\dfrac{\kappa_{1}^{ext} |A_{in}|^{2} g_{1}^{2} g_{2}^{2} (\omega_{m}^{2}+\frac{\gamma_{m}^{2}}{4})  }{(\alpha_{1R} \beta_{1R}-\alpha_{1I} \beta_{1I}+\gamma_{1R})^{2}+(\alpha_{1R} \beta_{1I}+\beta_{1R} \alpha_{1I}+\gamma_{1I})^{2}},
\end{equation}

where 

\begin{equation}
\alpha_{1R}= \dfrac{\kappa_{1} \gamma_{m}}{4} (1+C_{1}),   \alpha_{1I}=\dfrac{\kappa_{1} \omega_{m}}{2},
\end{equation}

\begin{equation}
\beta_{1R}=\frac{\kappa_{2} \gamma_{qd}}{4}(2+D_{2}), \\  \beta_{1I}=\frac{\Delta_{2} \gamma_{qd}}{2},
\end{equation}

\begin{equation}
\gamma_{1R}=\frac{D_{1} \kappa_{1} \gamma_{qd}}{4} (\frac{\kappa_{2} \gamma_{m}}{4}-\Delta_{2} \omega_{m}), \\ \gamma_{1I}= \frac{D_{1} \kappa_{1} \gamma_{qd}}{4} (\Delta_{2} \frac{\gamma_{m}}{2}+\frac{\omega_{m} \kappa_{2}}{2}).
\end{equation}

The subscript "s" indicates steady state.

\begin{figure}[ht]
\hspace{-0.7cm}
\begin{tabular}{cc}
\includegraphics [scale=0.60]{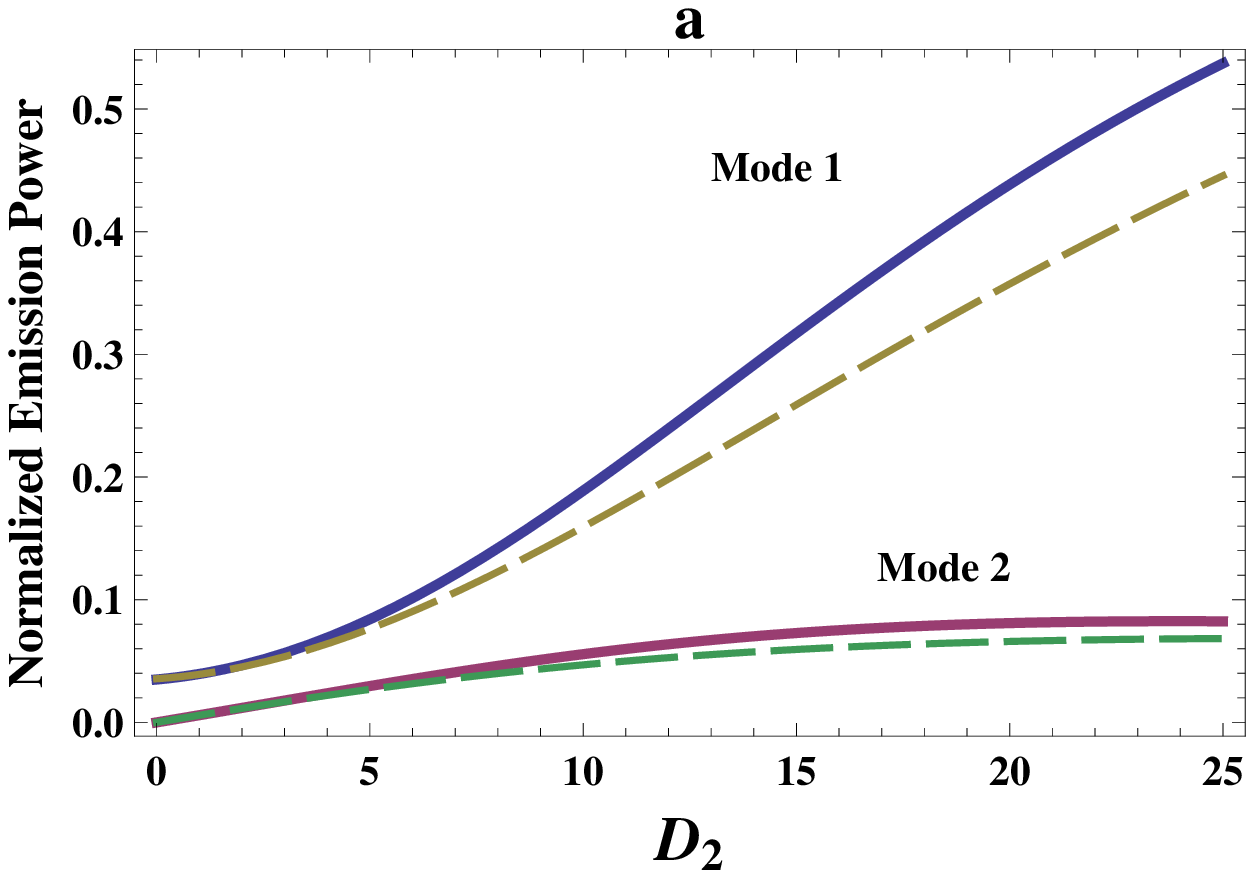} \includegraphics [scale=0.60] {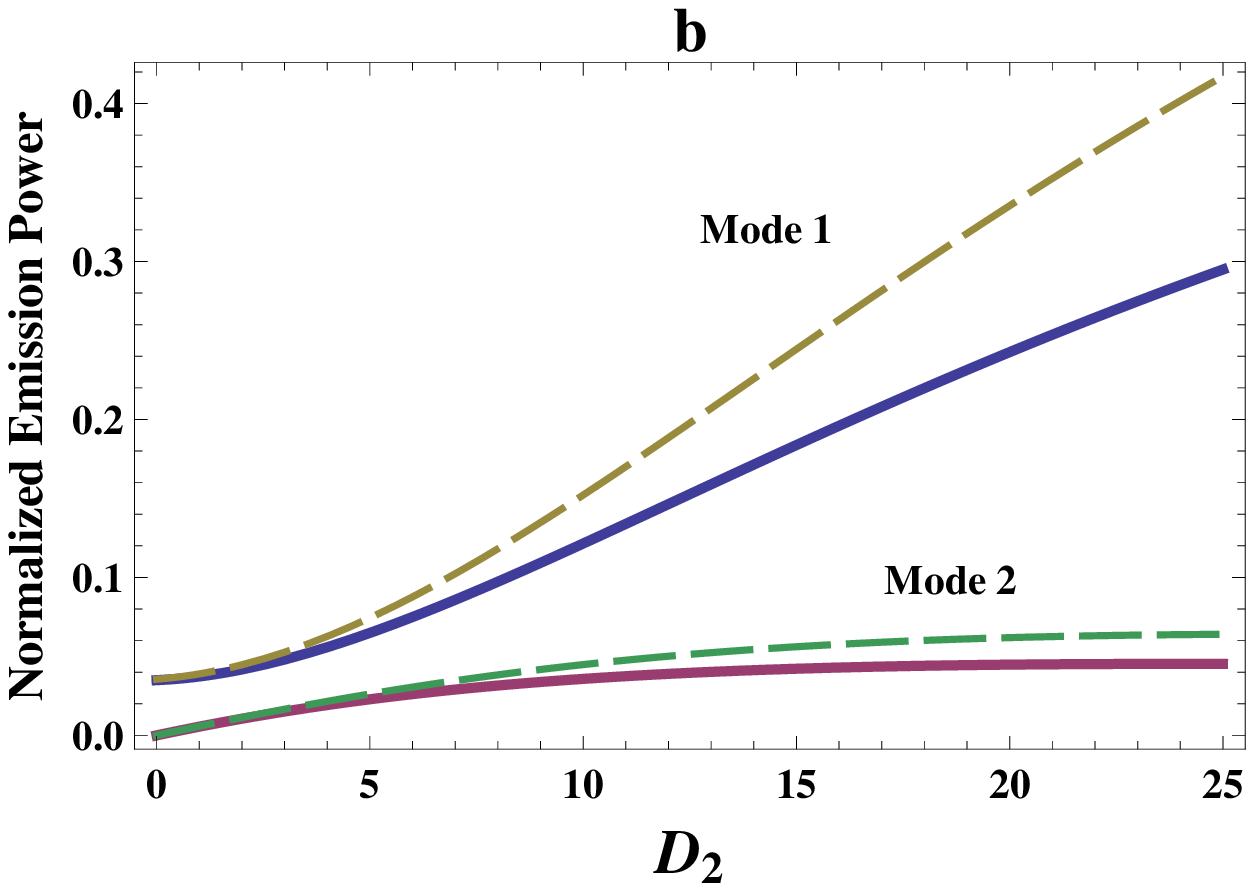}\\
 \end{tabular}
\caption{(Color online)  Normalized optical intracavity emission from mode 1 and mode 2 as a function of cavity cooperativity $D_{2}$ for the case $\Omega_{1} \neq 0$, $\Omega_{2}=0$ and for two different values of optomechanical cooperativity $C_{1}=1$ (dashed line), $C_{1}=10$ (solid line). (a) $D_{1}=4.3$, $\Delta_{1}=2.5 \kappa$, (b) $D_{1}=4.3$, $\Delta_{1}=-2.5 \kappa$. The other system parameters are chosen as $\gamma_{qd}=0.3 \kappa$, $\gamma_{m}=0.001 \kappa$, $\omega_{m}=0.01 \kappa$ and we have taken $\kappa_{1}=\kappa_{2}=\kappa$. } 
\label{fig:3}
\end{figure}

The corresponding steady state values of the intracavity modes for the case $\Omega_{2} \neq 0$, $\Omega_{1}=0$ is,

\begin{equation}
|a_{1s}|^{2}=\dfrac{\kappa_{1}^{ext} |A_{in}|^{2} (\alpha_{2R}^{2}+\alpha_{2I}^{2}) }{(\gamma_{2R} \delta_{2R}-\gamma_{2I} \delta_{2I}+\beta_{2R})^{2}+(\gamma_{2I} \delta_{2R}+\gamma_{2R} \delta_{2I}+\beta_{2I})^{2}},
\end{equation}

\begin{equation}
|a_{2s}|^{2}=\dfrac{\kappa_{1}^{ext} |A_{in}|^{2} g_{1}^{2} g_{2}^{2} (\omega_{m}^{2}+\frac{\gamma_{m}^{2}}{4})  }{(\gamma_{2R} \delta_{2R}-\gamma_{2I} \delta_{2I}+\beta_{2R})^{2}+(\gamma_{2I} \delta_{2R}+\gamma_{2R} \delta_{2I}+\beta_{2I})^{2}},
\end{equation}

where

\begin{equation}
\alpha_{2R}= \frac{\kappa_{2} \gamma_{qd} \gamma_{m}}{8} (1+C_{2}+D_{2}), \alpha_{2I}= \frac{\omega_{m} \kappa_{2} \gamma_{qd}}{4} (1+D_{2}),
\end{equation}

\begin{equation}
\beta_{2R}=\frac{D_{2} \kappa_{2} \gamma_{qd}}{4} \left ( \frac{\kappa_{1} \gamma_{m}}{4}-\Delta_{1} \omega_{m}  \right ), \beta_{2I}=\frac{D_{2} \kappa_{2} \gamma_{qd}}{4} \left ( \frac{\Delta_{1} \gamma_{m}}{2}+\frac{\omega_{m} \kappa_{1}}{2}\right),
\end{equation}

\begin{equation}
\delta_{2R}=\frac{\gamma_{m} \kappa_{2}}{4} (1+C_{2}), \delta_{2I}=\frac{\omega_{m} \kappa_{2}}{2},
\end{equation}

\begin{equation}
\gamma_{2R}= \frac{\kappa_{1} \gamma_{qd}}{4} (1+D_{1}), \gamma_{2I}=\frac{\Delta_{1} \gamma_{qd}}{2}.
\end{equation}

Here, $D_{j}=\frac{4 g_{j}^{2}}{\kappa_{j} \gamma_{qd}}$ is the cavity cooperativity for the optical mode $j$ and $C_{j}=\frac{4 \Omega_{j}^{2}}{\kappa_{j} \gamma_{m}}$ is the optomechanical cooperativity for the mechanical mode $j$.

Figure 2 shows the integrated emission intensity from mode 1 and mode 2 for the case $\Omega_{2} \neq 0$, $\Omega_{1}=0$. Fig. 2(a) and 2(b) corresponds to $\Delta_{1}=0.6 \kappa$ and $\Delta_{1}=- 0.6 \kappa$ respectively. Here we have taken $\kappa_{1}=\kappa_{2}=\kappa$. We observe that emission power  from mode 1 increases continuously with increasing $D_{2}$. The emission power from mode 2 increases with increasing $D_{2}$ but relatively in small amount. As $D_{2}$ increases, the emission power from mode 2 reaches a peak at $D_{2}= 3.0$ and then decreases continuosly. The effect of optomechanical cooperativity on power emission is also observed. For $\Delta_{1}=0.6 \kappa$, increasing $C_{2}$ from 1 (dashed line) to 10 (thick line), the emission power increases. The increase is relatively more for mode 1. This perhaphs occurs due to the fact that for $+ve$ detuning, the mechanical mode is transferring energy to the optical modes and this transfer of energy is more to mode 1. This process leads to cooling of the mechanical mode.
For $\Delta_{1}=-0.6 \kappa$ (Fig. 2(b)), increasing $C_{2}$ from 1 (dashed line) to 10 (thick line), the emission power decreases which indicates that energy is transferred from the optical modes to the mechanical mode and hence indicates heating of the mechanical mode. Fig.2(c) and 2(d) corresponds to the case when $D_{1}=6.3$. Analysis of Fig.2 reveals that on increasing the cavity cooperativity $D_{1}$ from 4.3 to 6.3, the optical emission from mode 2 exceeds that of mode 1 for low values of $D_{2}$ and subsequently decreases as $D_{2}$ increases.

Figure 3 displays the emission power when the mechanical mode couples with the optical mode 1 only ($\Omega_{2}=0$). It is observed that increase in emission power from mode 1 is much more rapid as compared to the previous case of $\Omega_{1}=0$. The increase in emission power from mode 2 is slow and reaches saturation much faster compared to Figure 2. The influence of optomechanical cooperativity $C_{1}$ on the emission power is same as discussed in for Fig.2 except for the fact that the influence of $C_{1}$ on mode 1 is much stronger.

\section{Mode conversion efficiency}

\begin{figure}[h]
\hspace{-1.3cm}
\begin{tabular}{cc}
\includegraphics [scale=0.60]{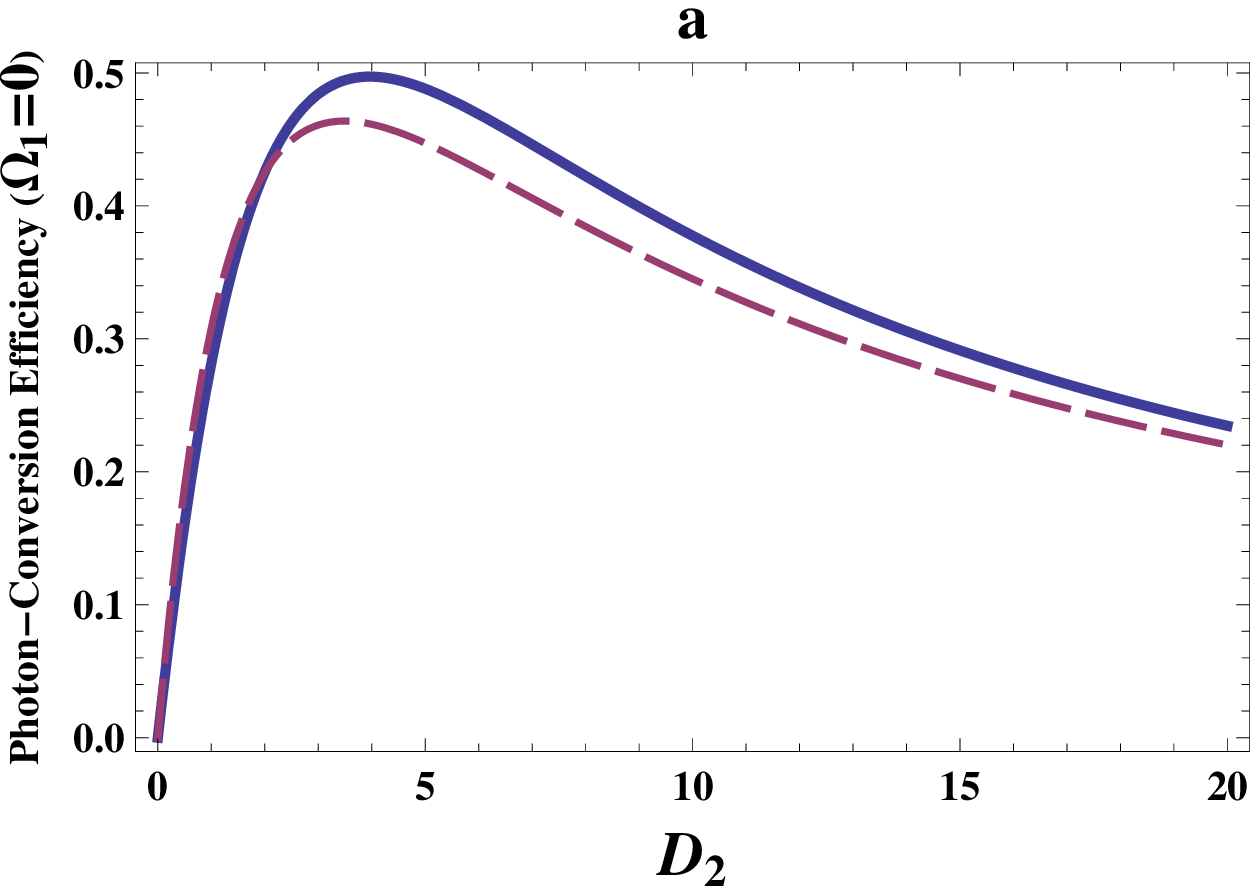} \hspace{4mm} \includegraphics [scale=0.60] {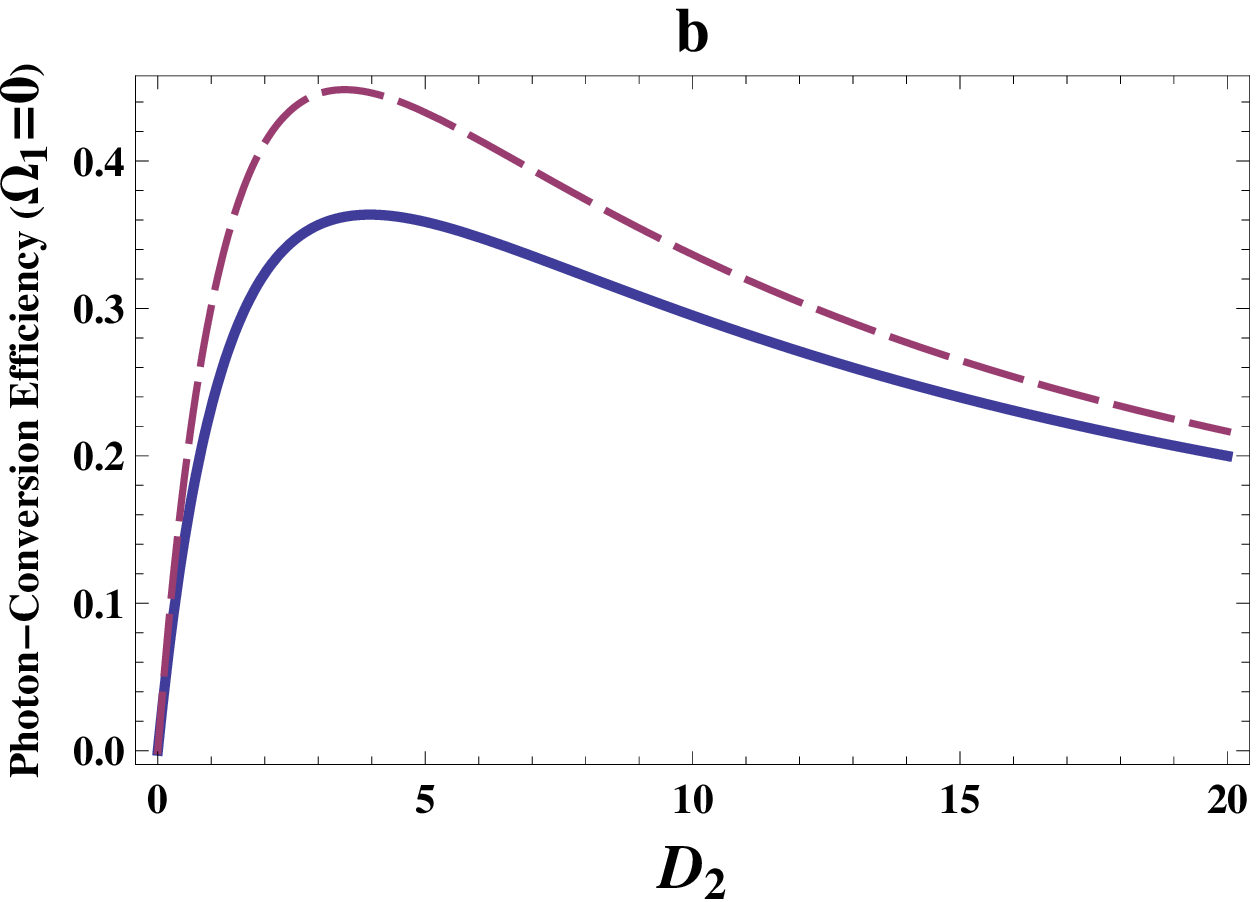}\\
\includegraphics [scale=0.60]{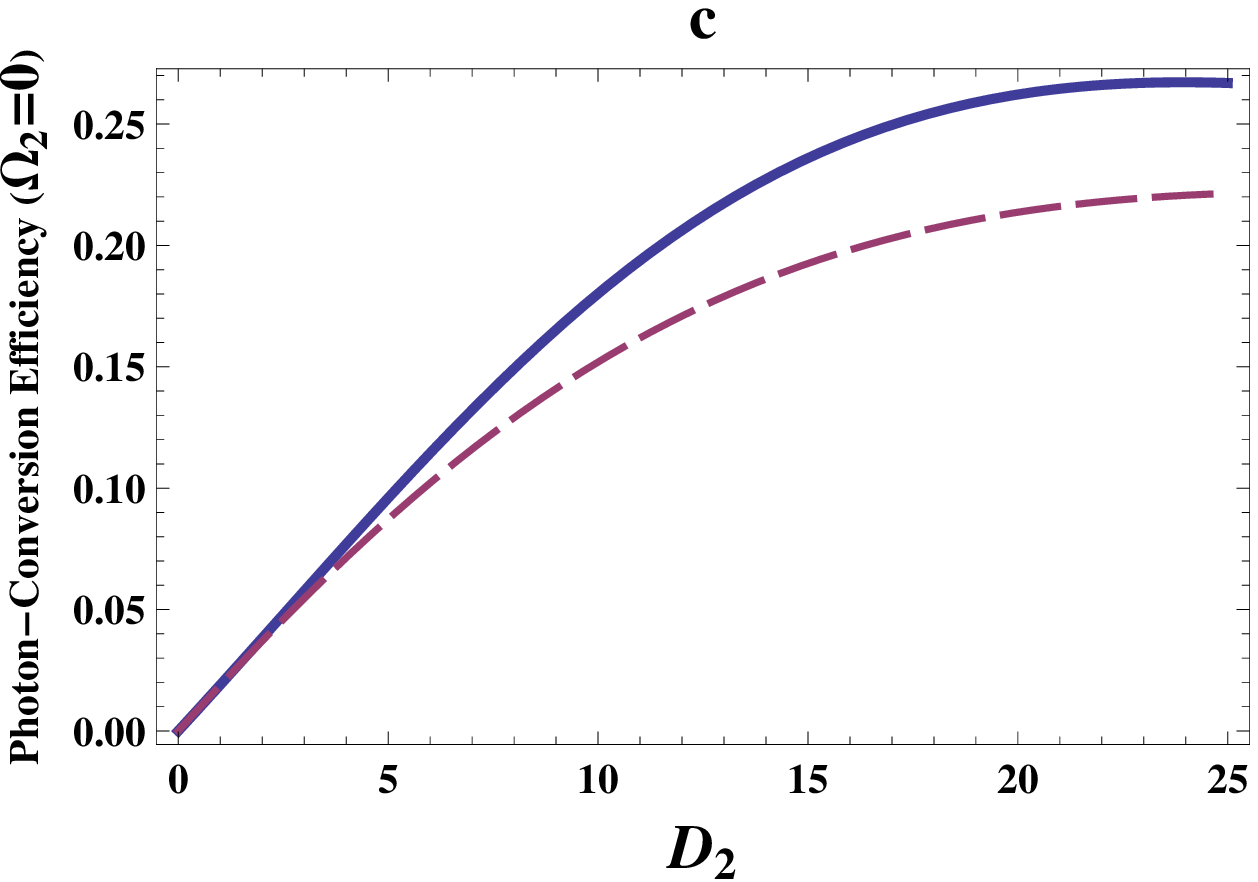} \hspace{4mm} \includegraphics [scale=0.60] {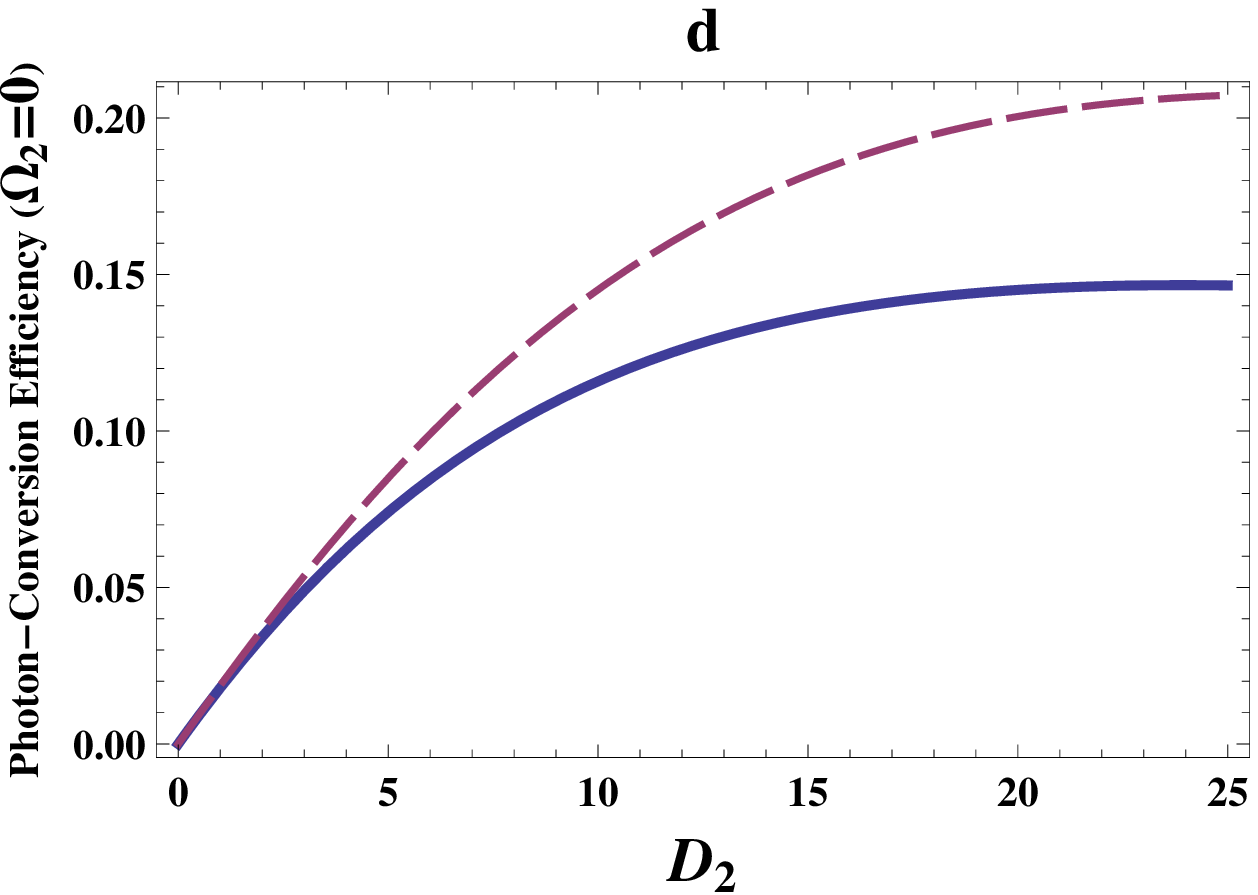}\\
\end{tabular}
\caption{(Color online) Photon mode conversion efficiency $\eta$ as a function of $D_{2}$ for the case $\Omega_{2} \neq 0$, $\Omega_{1}=0$ (plots (a) and (b)) and $\Omega_{1} \neq 0$, $\Omega_{2}=0$ (plots (c) and (d)). (a) $\Delta_{1}=0.6 \kappa$, (b) $\Delta_{1}= - 0.6 \kappa$, (c) $\Delta_{2}=2.5 \kappa$ and (d) $\Delta_{2}= - 2.5 \kappa$. The other system parameters are $\eta_{1}=\eta_{2}=0.9$, $D_{1}=4.3$, $\omega_{m}=0.01 \kappa$, $\gamma_{m}=0.001 \kappa$, $\gamma_{qd}=0.3 \kappa$ and $C_{2}=1$ (plots (a) and (b), dashed line), $C_{2}=10$ (plots (a) and (b), solid line),  $C_{1}=1$ (plots (c) and (d), dashed line), $C_{1}=10$ (plots (c) and (d), solid line). }
\label{fig:4}
\end{figure}

Let us now proceed to define and calculate the optical mode conversion efficiency ($\eta$) for our proposed system. Here $\eta$ is defined as, $\eta=\frac{I_{out}}{I_{in}}$, where $I_{in}=|A_{in}|^{2}$ is the input photon flux from mode 1 and $I_{out}= \kappa_{2}^{ext} |a_{2s}|^{2}$ is the output photon flux of mode 2. Here $\kappa_{i}^{ext} (i=1,2)$ is the effective output coupling rate of mode $i$. Thus the photon mode conversion efficiency for $\Omega_{2}=0$ case is written as,

\begin{equation}
\eta=\dfrac{\eta_{1} \eta_{2} \kappa_{1}^{2} \kappa_{2}^{2} \gamma_{qd}^{2} D_{1} D_{2} (\omega_{m}^{2}+\frac{\gamma_{m}^{2}}{4})}{16 (\alpha_{1R} \beta_{1R}-\alpha_{1I} \beta_{1I}+\gamma_{1R})^{2}+(\alpha_{1R} \beta_{1I}+\beta_{1R} \gamma_{1I}+\gamma_{1I})^{2}},
\end{equation}

where, $\eta_{i}=\frac{\kappa_{i}^{ext}}{\kappa_{i}} (i=1,2)$ is the output coupling ration. In a similar way, the photon mode conversion efficiency for $\Omega_{1}=0$ case is derived as,

\begin{equation}
\eta=\dfrac{\eta_{1} \eta_{2} \kappa_{1}^{2} \kappa_{2}^{2} \gamma_{qd}^{2} D_{1} D_{2} (\omega_{m}^{2}+\frac{\gamma_{m}^{2}}{4})}{16 (\gamma_{2R} \delta_{2R}-\gamma_{2I} \delta_{2I}+\beta_{2R})^{2}+(\gamma_{2I} \delta_{2R}+\gamma_{2R} \delta_{2I}+\beta_{2I})^{2}}.
\end{equation}

The optical mode conversion efficiency $\eta$ depends on the extend of destructive interference in the output of mode \citep{li}.

Figure 4 plots the photon mode-conversion efficiency as a function of $D_{2}$. Figure 4(a) and 4(b) displays the plot of $\eta$ for $\Omega_{1}=0$, $\Delta_{1}=0.6 \kappa$ and $\Delta_{1}=-0.6 \kappa$ respectively. On the other hand Fig. 4(c) and 4(d) shows the plot of $\eta$ for $\Omega_{2}=0$, $\Delta_{2}=2.5 \kappa$ and $\Delta_{2}=-2.5 \kappa$ respectively. In Fig.4(a), the photon mode conversion efficiency for $\Delta_{1}=0.6 \kappa$ increases as $D_{2}$ increases from $0$ to $4.3$. Beyond $D_{2}=4.3$, the efficiency decreases (i.e when $D_{2}> D_{1}$). The influence of optomechanical coupling for this case of $+ve$ detuning when going from $C_{2}=1$ to $C_{2}=10$ is visible only beyond $D_{1}=D_{2}$. A slight increase in efficiency is seen when $C_{2}=10$ as compared to $C_{2}=1$. For $\Delta_{1}=-0.6 \kappa$ (Fig.4(b)), increasing the optomechanical cooperativity from $C_{2}=1$ to $C_{2}=10$ significantly decreases the efficiency. In Fig.4(c) and 4(d), the photon-mode conversion efficiency is relatively less compared to Fig. 4(a) and 4(b) but the influence of optomechanical cooperativity on the photon-conversion efficiency is seen to be appreciably high.

\section{Dark and Bright modes}

In order to go deeper into the origin of optical mode conversion in our proposed setup, we introduce the concept of cavity dark and bright modes as

\bigskip

$\hat{a}_{B}=\dfrac{g_{1} \hat{a}_{1}+g_{2} \hat{a}_{2}}{\tilde{g}}$ and $\hat{a}_{D}=\dfrac{g_{2} \hat{a}_{1}-g_{1} \hat{a}_{2}}{\tilde{g}}$, where $\tilde{g}^{2}=g_{1}^{2}+g_{2}^{2}$. Making use of the above definition, we rewrite the Hamiltonian for $\Omega_{2}=0$, $\Omega_{1} \neq 0$ case with $\Delta_{1}=\Delta_{2}=\Delta_{d}=\Delta$ and $\kappa_{1}=\kappa_{2}=\kappa$ as,

\begin{eqnarray}
\tilde{H}_{1} &=& \hbar \Delta \hat{\sigma}^{\dagger} \hat{\sigma}+\hbar \Delta \hat{a}_{B}^{\dagger} \hat{a}_{B}+\hbar \Delta \hat{a}_{D}^{\dagger} \hat{a}_{D}+\hbar \omega_{m} \hat{b}^{\dagger} \hat{b}+\hbar \tilde{g} (\hat{a}_{B} \hat{\sigma}^{\dagger}+ \hat{a}_{B}^{\dagger} \hat{\sigma}) - \hbar \Omega_{1} \frac{g_{1}}{\tilde{g}} (\hat{a}_{B}^{\dagger} \hat{b}+\hat{a}_{B} \hat{b}^{\dagger}) \nonumber \\
 &-& \hbar \Omega_{1} \frac{g_{2}}{\tilde{g}} (\hat{a}_{D}^{\dagger} \hat{b}+\hat{a}_{D} \hat{b}^{\dagger})+\hbar \sqrt{\kappa_{1}^{ext}} \frac{g_{1}}{\tilde{g}}A_{in} ( \hat{a}_{B}+ \hat{a}_{B}^{\dagger})+\hbar \sqrt{\kappa_{1}^{ext}} \frac{g_{2}}{\tilde{g}} A_{in} ( \hat{a}_{D}+ \hat{a}_{D}^{\dagger}).
\end{eqnarray}

We clearly see that in Eqn.(20), there is no coupling between the QD and the dark mode, while the bright mode couples to the QD with an effective coupling $\tilde{g}$. On the other hand both the dark and bright modes couple to the mechanical mode with coupling $\Omega_{D}=\dfrac{\Omega_{1} g_{2}}{\tilde{g}}$ and $\Omega_{B}=\dfrac{\Omega_{1} g_{1}}{\tilde{g}}$ respectively. This indicates that optomechanics is appearing as a new handle to control the optical mode conversion using the QD-semiconductor cavity system. Using equations of motion for $\hat{a}_{B}$ and $\hat{a}_{D}$ from $\tilde{H}_{1}$, we calculate the steady state values of the bright and dark modes for $\Delta=0$ as,

\begin{equation}
a_{B}= F_{1} \sqrt{\frac{D_{1}}{D_{1}+D_{2}}}(i \omega_{m}+\frac{\gamma_{m}}{2}) ,
\end{equation}

\begin{equation}
a_{D}= F_{1} \sqrt{\frac{D_{2}}{D_{1}+D_{2}}}(i \omega_{m}+\frac{\gamma_{m}}{2}) (1+D_{1}+D_{2}) ,
\end{equation}

where 

\begin{equation}
F_{1}=\dfrac{-\frac{2i}{\kappa} \sqrt{\kappa_{1}^{ext}} A_{in} }{\frac{\gamma_{m}}{2} [(1+D_{1}+D_{2})+C_{1}(1+D_{2})]+i \omega_{m}(1+D_{1}+D_{2})}.
\end{equation}

Note that by putting $C_{1}=0$ (no optomechanics), we get the same expressions Eqns (24) and (25) of Li et. al \citep{li}. In order to explore the contributions of the dark and bright modes to the original cavity modes 1 and 2, we rewrite $a_{1}$ and $a_{2}$ in terms of $a_{B}$ and $a_{D}$ for $\Delta=0$ as,

\begin{equation}
a_{1}=F_{1} \dfrac{(i \omega_{m}+\frac{\gamma_{m}}{2}) (1+D_{1}+D_{2}) }{(D_{1}+D_{2})}   (D_{2}+\frac{D_{1}}{1+D_{1}+D_{2}}),
\end{equation}

\begin{equation}
a_{2}=F_{1} \dfrac{ \sqrt{D_{1} D_{2}} (1+D_{1}+D_{2}) }{(D_{1}+D_{2})}   (\frac{1}{1+D_{1}+D_{2}}-1) (i \omega_{m}+\frac{\gamma_{m}}{2}).
\end{equation}

As observed earlier \citep{li}, we distinctly find contributions of dark and bright modes to cavity mode 1 and mode 2. A distinct feature of our results is the contribution of the optomechanics apart from dark and bright modes in controlling the mode conversion. For $C_{1}=0$, our results reduces to those found in \citep{li}. We also introduce the dark-mode fraction $\chi$ defined as, $\chi=\dfrac{|a_{D}|^{2}}{|a_{D}|^{2}+|a_{B}|^{2}}$. This yields,

\begin{equation}
\chi=1- \frac{D_{1}}{D_{1}+D_{2}(1+D_{1}+D_{2})^{2}},
\end{equation}

which surprisingly is the same as that derived in ref \citep{li}. This indicates that for the case $\Omega_{2}=0$, $\Omega_{1} \neq 0$, the dark-mode fraction is uneffected by the presence of opto-mechanics. Proceeding in a similar manner, we now write the Hamiltonian for the $\Omega_{1}=0$, $\Omega_{2} \neq 0$ case as,

\begin{eqnarray}
\tilde{H}_{2} &=& \hbar \Delta \hat{\sigma}^{\dagger} \hat{\sigma}+\hbar \Delta \hat{a}_{B}^{\dagger} \hat{a}_{B}+\hbar \Delta \hat{a}_{D}^{\dagger} \hat{a}_{D}+\hbar \omega_{m} \hat{b}^{\dagger} \hat{b}+\hbar \tilde{g} (\hat{a}_{B} \hat{\sigma}^{\dagger}+ \hat{a}_{B}^{\dagger} \hat{\sigma}) - \hbar \Omega_{2} \frac{g_{2}}{\tilde{g}} (\hat{a}_{B}^{\dagger} \hat{b}+\hat{a}_{B} \hat{b}^{\dagger}) \nonumber \\
 &+& \hbar \Omega_{2} \frac{g_{1}}{\tilde{g}} (\hat{a}_{D}^{\dagger} \hat{b}+\hat{a}_{D} \hat{b}^{\dagger})+\hbar \sqrt{\kappa_{1}^{ext}} \frac{g_{1}}{\tilde{g}}A_{in} ( \hat{a}_{B}+ \hat{a}_{B}^{\dagger})+\hbar \sqrt{\kappa_{1}^{ext}} \frac{g_{2}}{\tilde{g}} A_{in} ( \hat{a}_{D}+ \hat{a}_{D}^{\dagger}).
\end{eqnarray}

One major difference between $\tilde{H}_{2}$ and $\tilde{H}_{1}$ that is visible is the change in the sign in the term that couples the dark mode with the mechanical mode. The corresponding expressions for the bright and dark mode amplitudes are,

\begin{equation}
a_{B}=F_{2} \sqrt{\frac{D_{1}}{D_{1}+D_{2}}} \left ( i \omega_{m}+\frac{\gamma_{m}}{2}(1+C_{2}) \right ),
\end{equation}

\begin{equation}
a_{D}=F_{2} \sqrt{\frac{D_{2}}{D_{1}+D_{2}}} \left ( \frac{\gamma_{m}}{2} (1+D_{1}+D_{2}+C_{2})+ i \omega_{m} (1+D_{1}+D_{2})  \right).
\end{equation}

where 

\begin{equation}
F_{2}=\dfrac{-\frac{2i}{\kappa} \sqrt{\kappa_{1}^{ext}} A_{in} }{\frac{\gamma_{m}}{2} [(1+D_{1}+D_{2})+C_{2}(1+D_{1})]+i \omega_{m}(1+D_{1}+D_{2})}.
\end{equation}

\begin{figure}[ht]
\hspace{-0.7cm}
\begin{tabular}{cc}
\includegraphics [scale=0.60]{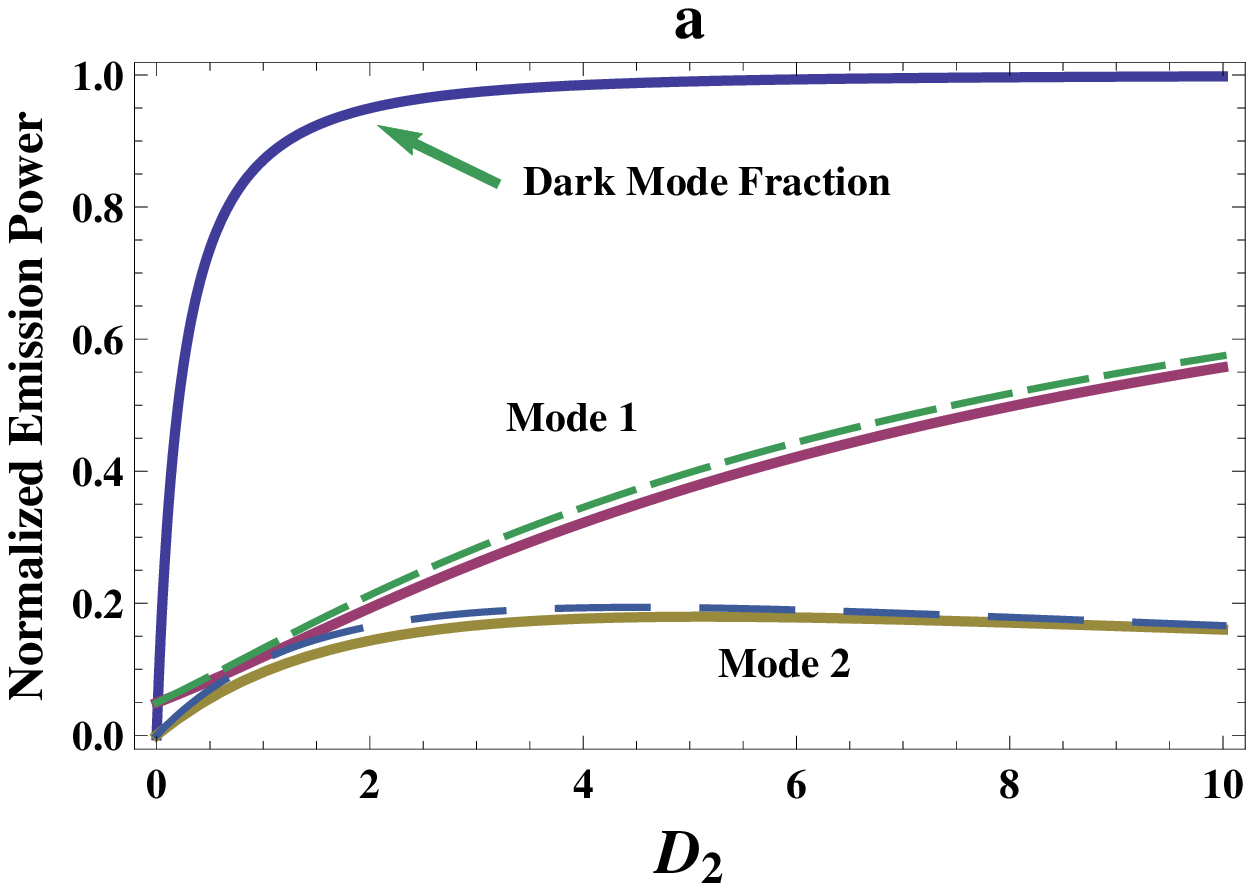} \includegraphics [scale=0.60] {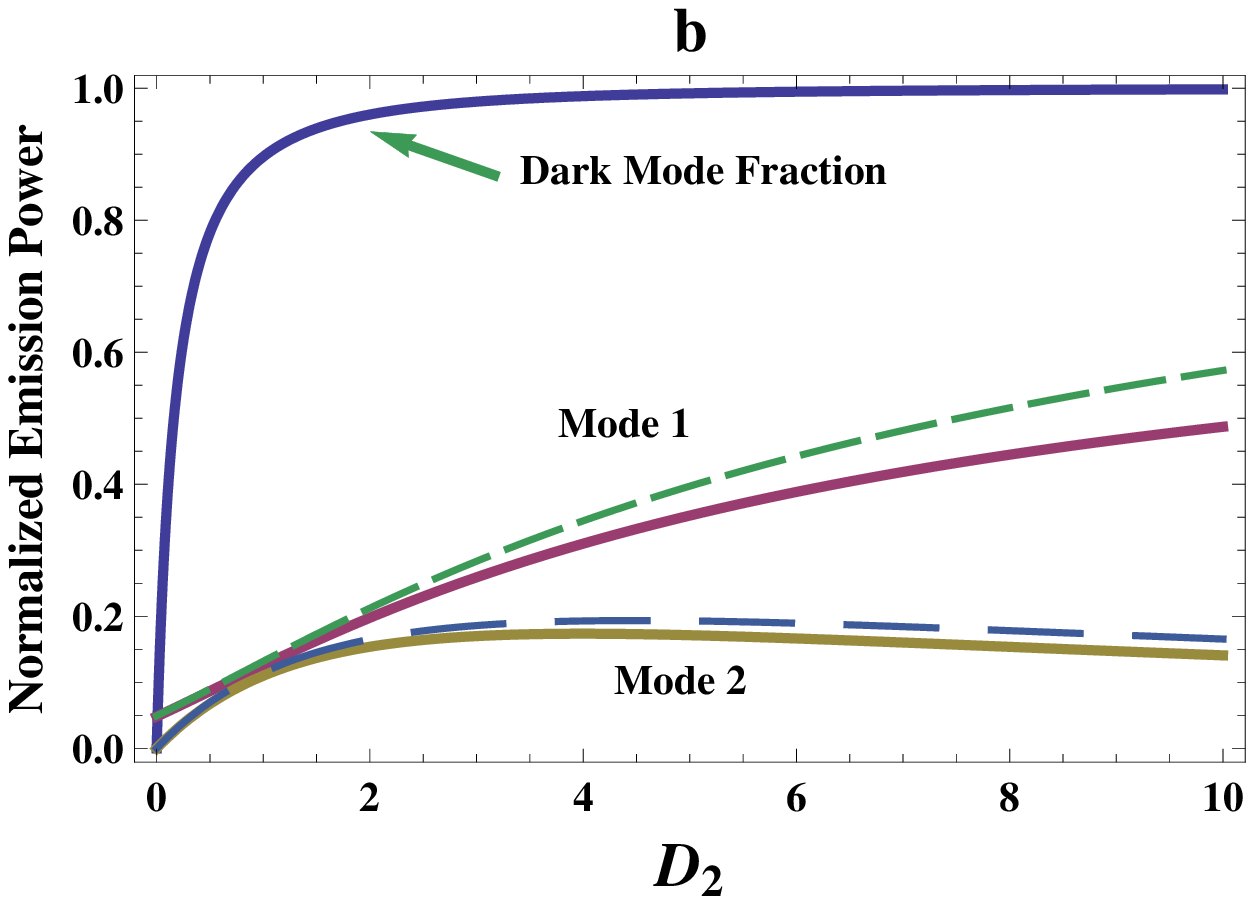}\\
\includegraphics [scale=0.60]{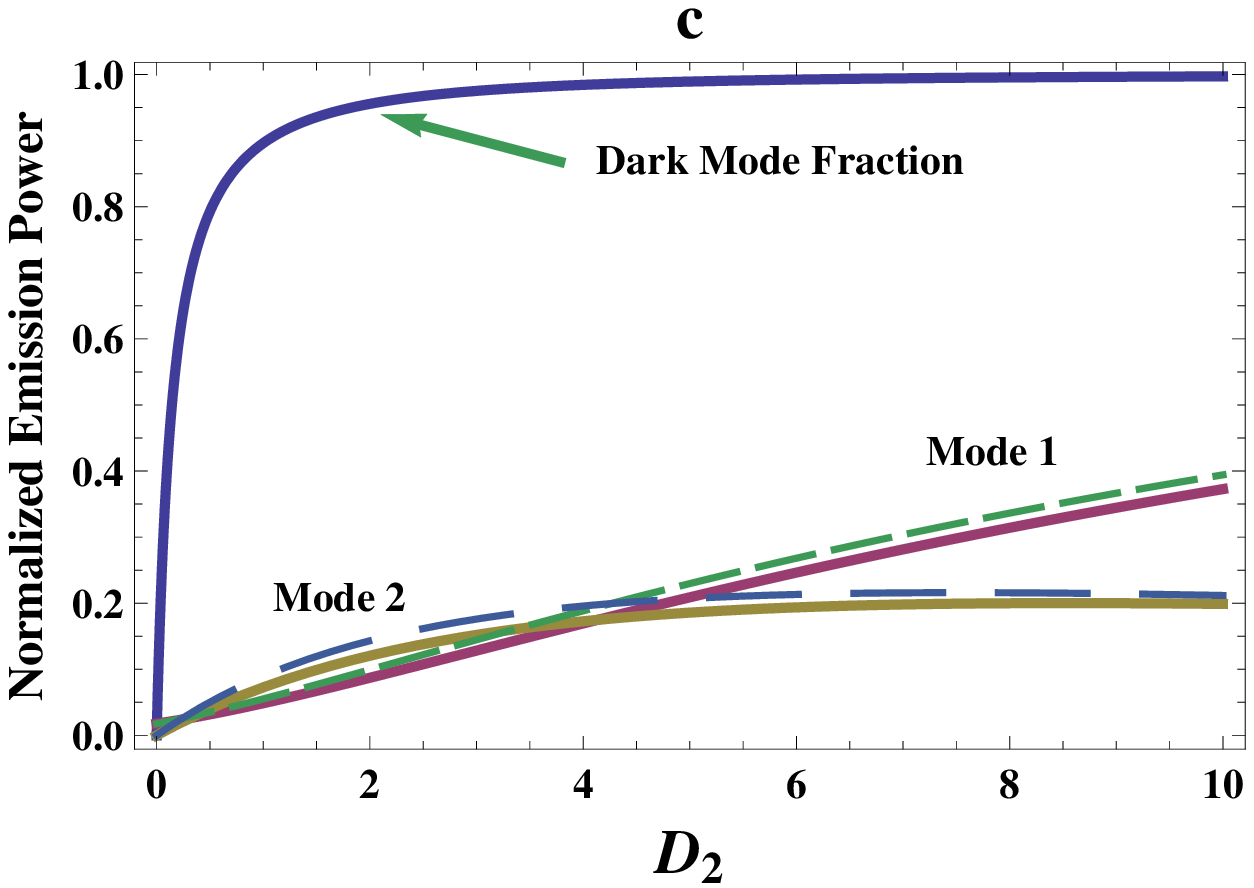} \includegraphics [scale=0.60] {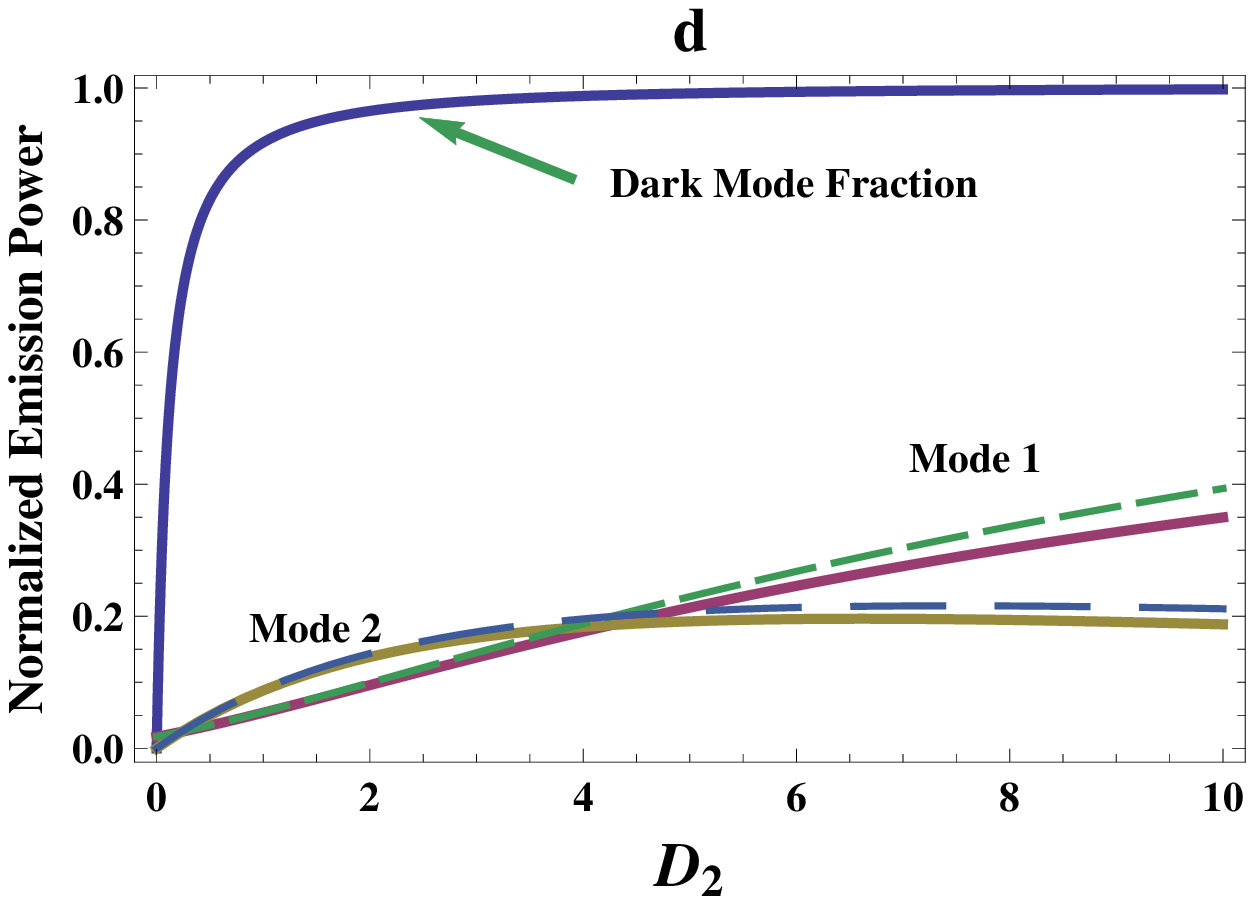}\\
 \end{tabular}
\caption{(Color online)   Dark-mode fraction $\chi$ and emission powers from mode 1 and mode 2 as a function of cavity cooperativity $D_{2}$. (a) $D_{1}=3.5$, $C_{2}=1,10$, (b) $D_{1}=3.5$, $C_{1}=1,10$, (c) $D_{1}=6.5$, $C_{2}=1,10$ and (d) $D_{1}=6.5$, $C_{1}=1,10$. The dashed line is for lower value of $C_{1} (C_{2})=1$ and the solid line is for the higher value of $C_{1} (C_{2})=10$ The other parameters are $\gamma_{m}=0.001 \kappa$ and $\omega_{m}= 0.01 \kappa$.} 
\label{fig:5}
\end{figure}

The modes 1 and 2 are rewritten in terms of $a_{B}$ and $a_{D}$ as,

\begin{equation}
a_{1}= F_{2} \dfrac{1+D_{1}+D_{2}}{D_{1}+D_{2}} \left ( \frac{\gamma_{m}}{2} \left [ D_{2} + \frac{D_{1}}{1+D_{1}+D_{2}} +\frac{C_{2} (D_{1}+D_{2})}{1+D_{1}+D_{2}}  \right] +i \omega_{m} \left [ D_{2} + \frac{D_{1}}{1+D_{1}+D_{2}} \right] \right)
\end{equation}

\begin{equation}
a_{2}=F_{2} \dfrac{ \sqrt{D_{1} D_{2}} (1+D_{1}+D_{2}) }{(D_{1}+D_{2})}   (\frac{1}{1+D_{1}+D_{2}}-1) (i \omega_{m}+\frac{\gamma_{m}}{2}).
\end{equation}

Compared to the previous case of $\Omega_{2}=0$, $\Omega_{1} \neq 0$, the expression for $a_{1}$ of Eqn. (31) is more complex with a non-trivial coupling with the mechanical mode. The expression for mode 2 does not change. The corresponding expression for the dark mode fraction turns out to be,

\begin{equation}
\chi= \dfrac{D_{2} \left [ \frac{\gamma_{m}^{2}}{4} (1+D_{1}+D_{2}+C_{2})^{2} + \omega_{m}^{2} (1+D_{1}+D_{2})^{2} \right] }{\frac{\gamma_{m}^{2}}{4} \left [ D_{2} (1+D_{1}+D_{2}+C_{2})^{2}+D_{1} (1+C_{2})^{2} \right]+ \omega_{m}^{2} \left [ D_{2}(1+D_{1}+D_{2})^{2}+D_{1} \right]}.
\end{equation}

From the expression of cavity mode 2, it is clear that there is a destructive interference between the bright and dark modes and hence 2 can be excited by suppressing the bright mode. For the $\Omega_{2}=0$ case, bright mode can be suppressed by keeping $D_{1}=D_{2}>>1$. On the other hand for $\Omega_{1}=0$ case, the optomechanical cooperativity $C_{2}$ now is very convenient handle to control the suppression of the bright mode. Fig.5 plots the calculated dark-mode fraction and the mode 1 and mode 2 emission power as a function of $D_{2}$. As before, we notice that when $D_{1}=6.5$, the emission power of mode 2 is higher than that of mode 1 for $D_{2}<4.0$. For increasing $D_{2}$, the emission power of mode 2 gradually saturates and becomes less compared to that of mode 1. Also  observed is the fact that the system is driven towards dark mode as $D_{2}$ increases. 

\section{Conclusion}

In summary, we have proposed a hybrid semiconductor microcavity system for coherently controlling the conversion between two cavity optical modes using a two-level quantum dot and optomechanics. The composite hybrid system can be experimentally fabricated using distributed Bragg reflectors and resonantly driven with a CW laser. We consider two specific cases in which the mechanical mode is selectively coupled to one of the optical modes. We have shown analytically that for specific system parameters, increasing or decreasing the optomechanical cooperativity tunes the efficiency of conversion between the two optical modes. In particular, we find that the conversion efficiency is higher when the mechanical mode is coupled to the undriven optical cavity mode. We also demonstrate that the mechanical mode can control the enhancement of the dark mode and suppression of the bright mode which leads to the desired optical mode conversion. This observation originates from the fact that the dark mode does not couple to the quantum dot while the mechanical mode couples to both the dark and the bright mode. 

\section{Acknowlegements}
A. B. B acknowledges Birla Institute of Technology and Science , Pilani for the facilities to carry out this research. A.B.B is also thankful to SERB-Department of Science and Technology, New Delhi for the financial support under SERB Project No. : EMR/ 2017/0019870. S.A would like to thank the School of Physical Sciences, JNU for their support.

\end{document}